\documentclass[aps,twocolumn,showpacs,groupedaddress,superscriptaddress,amsmath,amssymb,prb]{revtex4}
\usepackage{epsfig}

\begin{document}

\title{How square ice helps lubrication}

\author{Astrid S. de Wijn}\affiliation{Department of Physics, Stockholm University, 10691 Stockholm, Sweden}
\affiliation{Department of Engineering Design and materials, Norwegian University of Science and Technology, 7491 Trondheim, Norway}
\author{Lars G.~M.~Pettersson}
\affiliation{Department of Physics, Stockholm University, 10691 Stockholm, Sweden}

\begin{abstract}
In the context of friction we use atomistic molecular-dynamics simulations to investigate water confined between graphene sheets over a wide range of pressures.
We find that thermal equilibration of the confined water is hindered at high pressures.
We demonstrate that, under the right conditions, square ice can form in an asperity, and that it is similar to cubic ice VII and ice X.
We simulate sliding of atomically flat graphite on the square ice and find extremely low friction due to structural superlubricity.
The conditions needed for square ice to form correspond to low sliding speeds, and we suggest that the ice observed in experiments of friction on wet graphite is of this type.

\end{abstract}

\maketitle

\section{Introduction}
The combination of water with graphite or graphene is under active investigation in several fields for a number of reasons.
In the field of tribology, it is of interest due to the action of graphite powder as a solid lubricant, which is far more effective under humid conditions than in vacuum or dry air.
This is opposite to the case for other solid lubricants, such as WS$_2$ and MoS$_2$~\cite{solidlubricantcoatingsreview}.
Moreover, water alone is a poor lubricant, due to its low viscosity-pressure coefficient.
While suggestions have been made as to the reason behind water's beneficial effects on graphite as a lubricant~\cite{onssolidlubricant,graphenenanoscrolls}, this effect is not yet understood.

Under sufficiently strong confinement water doesn't crystallize at ambient pressure and thus, by confining water inside, e.g., nanotubes or zeolite pores it is possible to investigate liquid water well below the temperature of homogeneous ice nucleation (for a recent review, see~\cite{ChemRev2016}). Water confined between hydrophobic surfaces exhibits a complex behavior.
A number of high-profile experimental~\cite{squareicegeimnature,grapheneonmica,Severin2012,verdaguer2013,kimmel2009,vilhena2016} and numerical~\cite{chen2diceprl,watergraphenesims,watergraphenesimsDFT,watergraphenesimsstructurelessplates,kimmel2009,vilhena2016,giovambattista2009,tocci2014,giovambattista2009} studies have investigated water confined using graphene or carbon nanotubes under various conditions.
These studies have found that when water is confined between graphene sheets or graphite, there is clear structure in the direction normal to the surfaces but the in-plane order is typically either liquid-like or hexagonal.
In the case of the hexagonal structure there are, however, different opinions on whether or not it is related to the graphite structure~\cite{watergraphenesims,watergraphenesimsstructurelessplates}. Confinement is, however, not required for the hexagonal structure to develop under certain conditions. Kimmel et al.~\cite{kimmel2009} found that water deposited on graphene on Pt(111) at 100--130 ~K under ultra-high vacuum conditions forms a two-layer hexagonal ice structure. The structure was determined using LEED and RAIRS and in conjunction with {\it ab initio} molecular dynamics (MD) simulations a fully hydrogen-bonded two-layer hexagonal ice structure was concluded with no hydrogens pointing towards the graphene.
In simulations of water in carbon nanotubes~\cite{nanotubesquareice}, another structure was found where water arranges into a rolled-up square structure.
Square ice has also been observed in experimental studies of water confined between graphene sheets~\cite{squareicegeimnature}.
Han et al. \cite{watergraphenesimsstructurelessplates} focused on the question whether a critical point could exist on the liquid-solid coexistence line and used the pressure dependence of water structure confined between two smooth hydrophobic plates to derive the phase diagram in this region.
They find only a narrow coexistence region between the liquid and square ice solid without commenting on the structure of the square ice. An extensive study of the structure of a monolayer of ice between graphene sheets as function of the applied pressure was reported by Chen et al.~\cite{chen2diceprl}. Under strong confinement a flat square ice was found to be the most stable form in the pressure range 2--4~GPa.

Different aspects on the role of water in lubrication have been reported~\cite{jinesh2006,graphenenanoscrolls,chenstickslipcontrol,vilhena2016,tocci2014}.
In their experiments Jinesh and Frenken found capillary condensation of water, which at room temperature formed ice between the asperity and substrate, to lead to sticking rather than lubrication~\cite{jinesh2006}.
Similarly, Berman et al.~\cite{graphenenanoscrolls} found that in simulations water increased friction in their system where graphene flakes were combined with nanodiamonds.
Chen et al.~\cite{chenstickslipcontrol} studied the effect of water between graphene sheets at liquid conditions (low pressures) and found an absense of stick-slip dynamics. Vilhena and coworkers~\cite{vilhena2016} simulated an AFM diamond tip on graphene with one water layer between the tip and substrate and found significantly reduced friction. 
In the present work we find that water in between graphene sheets under the right sliding conditions, in the form of room temperature square ice, leads to superlubricity.

All of the numerical studies mentioned above deal either with the equilibrium phase diagram of confined water or sliding at constant and relatively low pressures.
However, a sliding contact produces local changes in pressure and temperature and is typically out of equilibrium.
Moreover, the structure in a sliding contact can be of enormous influence on the friction.  Mismatched lattice parameters, especially, can lead to extremely low friction, known as structural superlubricity (see, for example~\cite{Dienwiebel2004}).

Experiments measuring the friction of an AFM tip on graphite covered in water~\cite{jinesh2006,jinesh2008} have indeed demonstrated interesting velocity dependence.
At low sliding velocities, a stick-slip period of 3.8~\AA{} was observed, which does not correspond to any period of graphite.
At higher sliding speeds, the normal graphite period was present instead.  This behavior was also found to be related to air humidity.
This implies that under at least some sliding conditions, ice can be formed on a graphite substrate at room temperature.

In this work, we use MD simulations to study the combination of water and graphite in the context of friction and out-of-equilibrium dynamics.
As the pressure in a single asperity can be up to several tens of GPa, we pay particular attention to high pressures.
We find that thermal equilibration of the confined water is hindered at high pressures.
We demonstrate that, under the right conditions, square ice can form, and that it is similar to cubic ice VII and ice X.
The conditions needed for this equilibration correspond to low sliding speeds, and therefore suggest that the ice observed in the friction experiments~\cite{jinesh2006,jinesh2008}, is cubic VII or X.

\section{Description of the simulations}

Surfaces are generally not flat, and the actual area of contact between two sliding surfaces consists of a large number of microscopic contact asperities, where the peaks in the roughness meet.
For reasons of computational power, it is impossible to simulate two realistic surfaces with a number of contact asperities.
Even a single asperity moving at the velocities of an AFM experiment is not yet possible~\cite{lowspeedfrictionsimulations}.
We therefore investigate a relatively small single asperity under a range of conditions that are likely to occur in a large sliding system.

\begin{figure}
\epsfig{figure=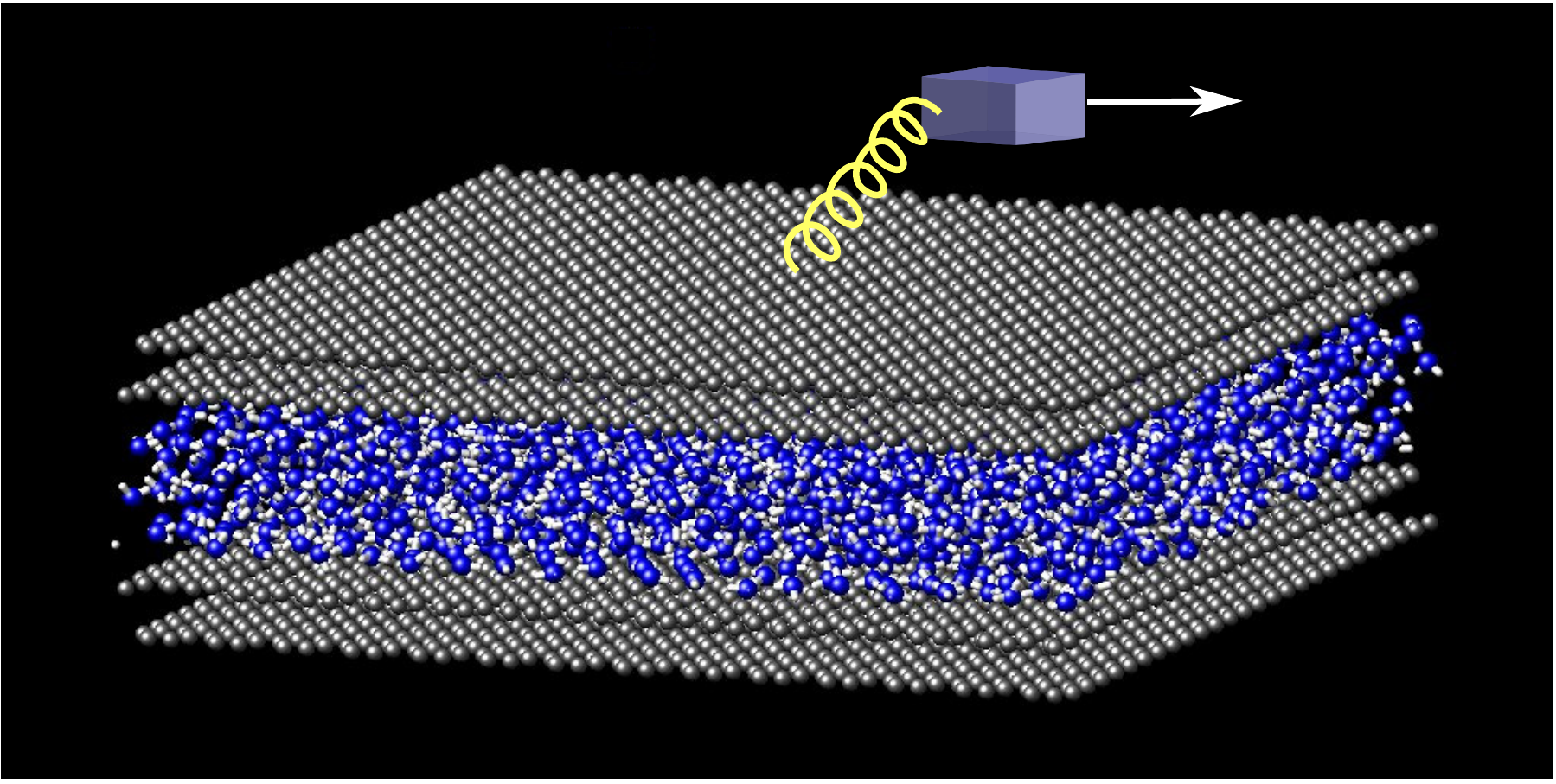,width=8.6cm}
\caption{\label{fig:simulationsetup}A sketch of our simulation box.  Carbon atoms are shown in grey, oxygen in blue, and hydrogen in white.
The topmost and bottommost graphene sheets are rigid.  The topmost sheet is attached with a spring to a support that moves at constant velocity during sliding runs.
}
\end{figure}

We simulate various amounts of water molecules between two bi-layer graphene slabs as illustrated in Fig.~\ref{fig:simulationsetup} with a total of $4N_\mathrm{C}=5824$ atoms where $N_\mathrm{C}$ is the number of carbon atoms in each graphene sheet in the periodically repeated simulation cell.
The simulations were performed using the MD package LAMMPS~\cite{lammps} with water described by the TIP4P/2005 model~\cite{TIP4P-2005}, which gives a very good overall description of bulk water and the ices, including at high pressures~\cite{TIP4P-2005,aragones2009}. The graphene sheets were described using the AIREBO reactive force-field~\cite{airebo}.
The interaction between water molecules and graphene is modeled with a Lennard-Jones potential between the oxygen and carbon atoms with parameters $\epsilon=4.063$~meV, $\sigma=0.319$~nm~\cite{watergraphenesimsDFT}. We note that, as also found experimentally~\cite{kimmel2009}, the water-water interaction dominates over the interaction between water and graphene.
The carbon, oxygen, and hydrogen atoms are placed in a simulation box with periodic boundary conditions in $x$ and $y$ with sizes of 6.395~nm by 5.964~nm.  In the $z$ direction (orthogonal to the slab) the box is so large as to be effectively infinite (800~nm).

The outer carbon layers are kept internally rigid while the middle inner layers are mobile and internally fully flexible.
The AIREBO potential, like most carbon potentials with long-range interactions based on Lennard-Jones~\cite{graphenecorrugationproblem}, has a problem with underestimating the inter-layer corrugation.
In order to prevent the mobile layers from slipping relative to their nearby rigid layers, especially during the sliding simulations, we add a spring between the centers of mass of each mobile layer and its nearest rigid layer.
The spring constant is equal to 41~meV/\AA{}$^2$ per atom, which was chosen so as to be consistent with experimentally determined inter-layer interactions for small displacements.
The bottommost layer is kept at a fixed position while the topmost layer is coupled in the $x$ and $y$ directions with a spring (spring constant also 41~meV/\AA{}$^2$ per atom) to a support that is kept at a fixed position during equilibration, and moves at constant velocity during sliding.
Pressure is applied through a uniformly distributed force on the topmost carbon layer.
We also vary the amount of water in our simulation between $N_\mathrm{O}/N_\mathrm{C} = 0.16$ and $1.33$, where we denote the total number of water molecules by $N_\mathrm{O}$.

In a normal equilibrium simulation of a confined liquid, it would be the pressure that determines $N_\mathrm{O}$, through a reservoir with a barostat.
When the load on the contact is increased, the liquid flows out.  However, in reality, when the surfaces are very large or rough, this squeezing out of the liquid can be hindered~\cite{perssonboek}.
Thus, in a large moving contact, it is not a given that the system is in such equilibrium.
For our purposes, there is therefore no direct link between the load/pressure and $N_\mathrm{O}$, and these must be treated as independent parameters.
While this means that in our simulations we must investigate a larger set of parameter combinations, we do not need to simulate a large liquid reservoir.

We employ Langevin dynamics for thermostatting with damping constant chosen equal to 1 (ps)$^{-1}$.
This damping coefficient is sufficiently small so as to not cause severe distortions of the dynamics even in the areas where it is applied~\cite{merel}.
The thermostat is always applied to all mobile carbon atoms and, whenever we do not slide the system, the water molecules are also thermostatted.
In friction simulations, however, care should always be taken with thermostatting, as this can severely distort results~\cite{merel}.
Moreover, in a sliding system in an experiment, heat must be removed through
the surfaces, in this case the carbon atoms.
Whenever we subject the system to sliding, we therefore only thermostat the mobile carbon atoms.
While we have chosen this thermostatting in order to be careful, we do not believe that thermostatting the water would have made any difference to our results, as there was very little change in temperature of the water or carbons.

Initial conditions are constructed by arranging the desired number of water molecules in a regular grid (but not an equilibrium structure) at approximately the density of water under ambient conditions.
The carbon slabs are placed at initial positions far enough apart to contain the water molecules.
We then simulate the system for some time under various load and temperature conditions as described below.

\subsection{Equillibration procedures}
Since we are interested in studying friction, which is a nonequilibrium process, we must investigate also how equilibrium can be reached.
Specifically, a moving asperity leads to time-dependent load variations which can affect the structure of the confined material
We thus first investigate the equilibration of our system under two different sequences of conditions, one that should produce the true equilibrium and one that mimics the conditions during sliding.
In equilibration method one (temperature-scan method), we start immediately by applying high pressure, but also high temperature (2000~K), and then slowly reduce the temperature to 293~K over a time interval of 5~ns by ramping it down linearly; this corresponds to simulated annealing.
In equilibration method two (load-scan method), the system starts from room temperature, but the pressure is increased in steps at intervals of 50~ps to 0.01, 0.02, 0.1, nN/atom, and after that it is incremented by 0.1 nN/atom every 50~ps.
 
These two different equilibration methods give an indication of what happens between real materials under different sliding conditions.
The load must be carried entirely by tiny asperities that are always present on the nanoscale, and as a result the local pressures can reach up to many GPa.
Moreover, as the asperities are moving, the pressure can vary rapidly as an asperity passes by.
At high sliding speeds, the pressure in the front of the contact increases rapidly when the asperity approaches.
This condition is best represented by the load-scan method.
At low sliding speeds, the pressure changes more slowly, and the confined water has ample time to equilibrate properly.  The water in the contact is thus more likely to reach the global equilibrium, which is more easily reached in simulations by the temperature-scan method.

The temperature-scan method is known to reliably provide the equilibrium structure of bulk water at room temperature under high pressure, while the load-scan method does not.
Specifically bulk ice VII is difficult to obtain in simulations by ramping up the pressure, but can be easily found by starting from high temperature and pressure and cooling the simulated system down slowly.

\section{Results}

\begin{figure}
\medskip
~~~~~~~(a)\hfill\strut\\[-5ex]
\epsfig{figure=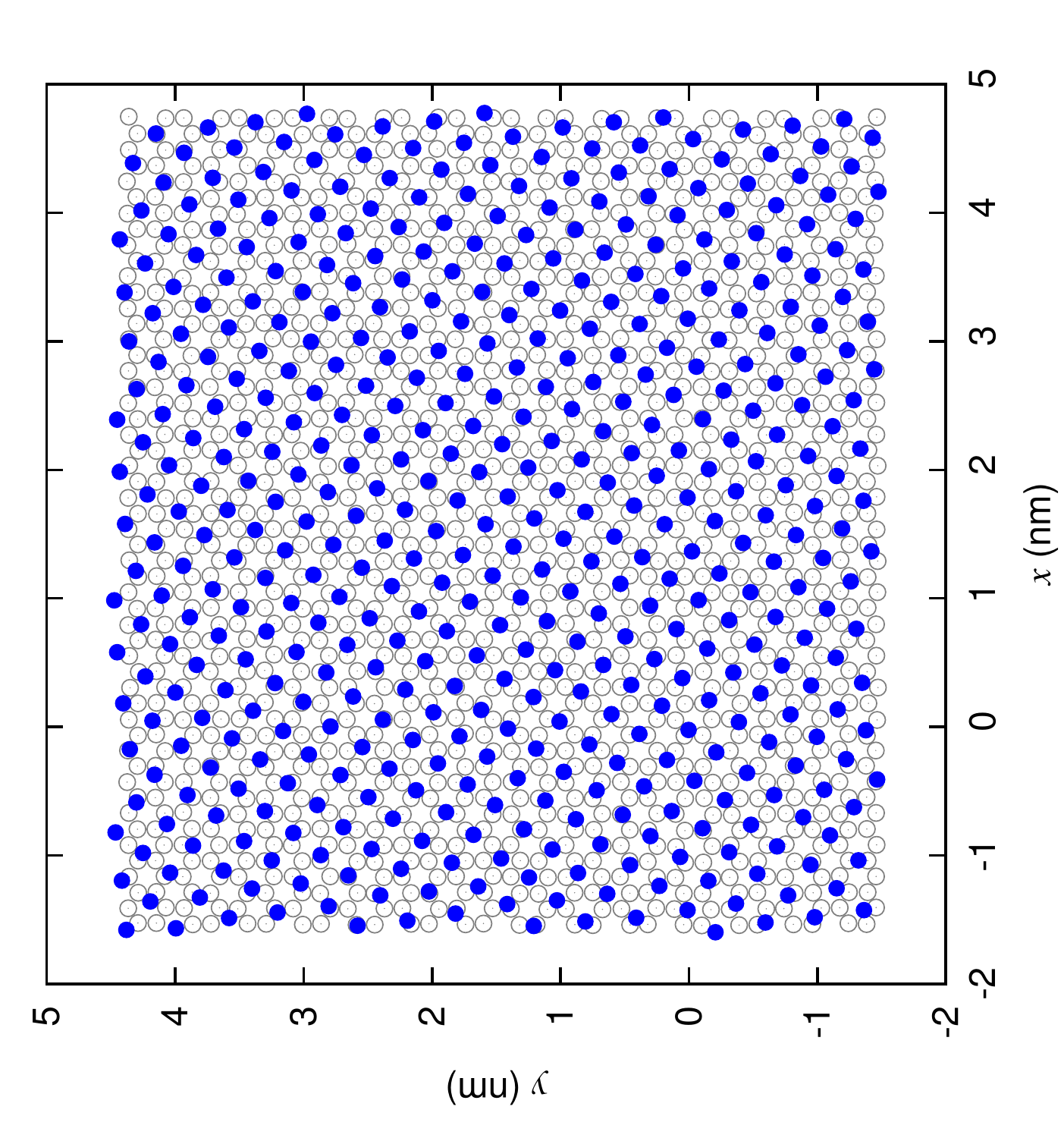,angle=270,width=6.6cm}\\
\medskip
~~~~~~~(b)\hfill\strut\\[-5ex]
\epsfig{figure=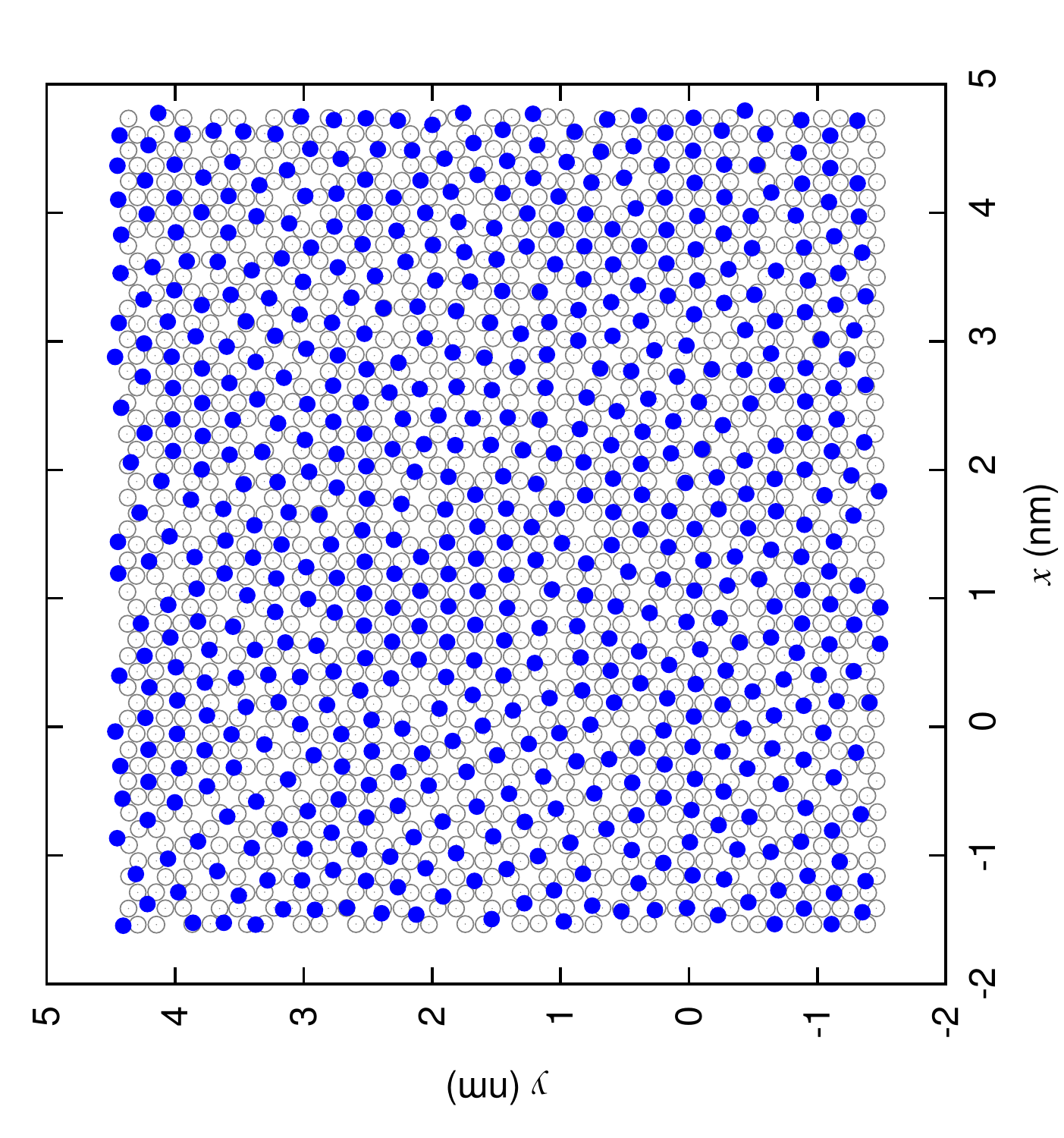,angle=270,width=6.6cm}\\
\medskip
~~~~~~~(c)\hfill\strut\\[-5ex]
\epsfig{figure=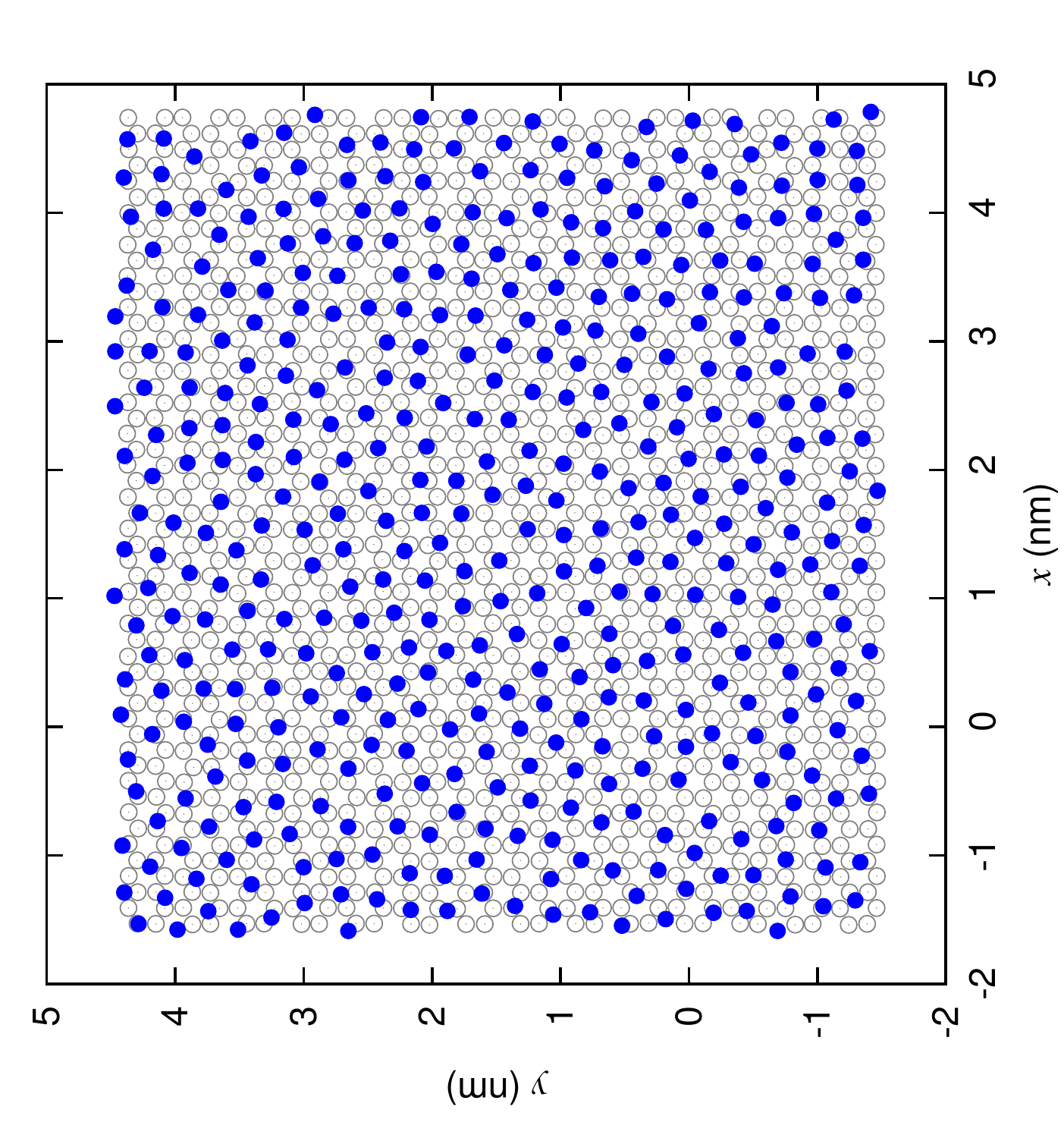,angle=270,width=6.6cm}
\caption{
The oxygens in the bottom layer of the water (filled symbols) together with the carbons in the top layer of the bottom graphene slab (open symbols) for three different cases: ice-like obtained by slowly reducing the temperature at high pressure (temperature-scan, top), liquid-like obtained by rapidly increasing the pressure at constant temperature (load-scan, middle) and liquid-like at lower pressure (bottom).  At low loads the system equilibrates more easily.
Ice-like structures appear at sufficiently high pressures, after careful equilibration, as expected from the phase diagram of bulk water.
In these simulations $N_\mathrm{O}/N_\mathrm{C}=1.33$ and the loads were (ramped up to) 1.0~nN/atom for (a) and (b), and 0.1~nN/atom for (c).
\label{fig:bottomlayers}
}
\end{figure}

At low pressures/loads, the water remains liquid for both equilibration methods.
At higher loads, however, the resulting structure depends on how the system approaches equilibrium.

We illustrate the different structures in Fig.~\ref{fig:bottomlayers}.
At high load, we find that the temperature-scan method produces a crystalline structure (Fig.~\ref{fig:bottomlayers}a).
The O atoms in the second layer (not shown) sit on top of the empty sites in the middle of the squares in the first layer.  The atoms of the third layer sit directly above those in the first layer.  This clearly indicates a BCC structure.
The load-scan method on the other hand does not produce a crystalline structure, but rather something more reminiscent of the liquid structures at low load.  Due to the high load, however, the arrangement of the water molecules follows somewhat the symmetry of the graphene sheets.

The true global minimum is the crystalline structure found from the temperature-scan method, which corresponds to annealing the sample.
The high-pressure confinement of the load-scan method is more prone to lead to a local minimum and thus more likely a non-crystalline structure. In fact, we find that the final structure in this case depends on the rate at which the pressure is increased where
a slower ramping up of the load tends to lead to a more crystalline, less amorphous structure.
We have also tested very fast ramping up, which inhibits crystallization even more.
In our simulations, we were never able to obtain anything as crystalline from a slow load ramp as we found from the temperature scan method.
Even if the system was simulated for a long time after ramping up the load, it still did not reach the true, crystalline, equilibrium.
This is consistent with what is known about ice structures at room temperature and high pressures,where, e.g.,  ice VII can most easily be obtained by cooling down water under high pressure.

In the context of MD simulations of water interacting with graphene, this is a crucial difference.  MD simulations of this confined system~\cite{watergraphenesims,watergraphenesimsstructurelessplates} so far have employed methods similar to the load-scan method.
Thus it is not surprising that these simulations were not able to consistently reproduce the square/cubic structure of ice VII (or X) that has been seen in experiments~\cite{squareicegeimnature}.

\section{Classification of structures}

\begin{figure}
~~~(a)\hfill\strut\\[-5ex]
\epsfig{figure=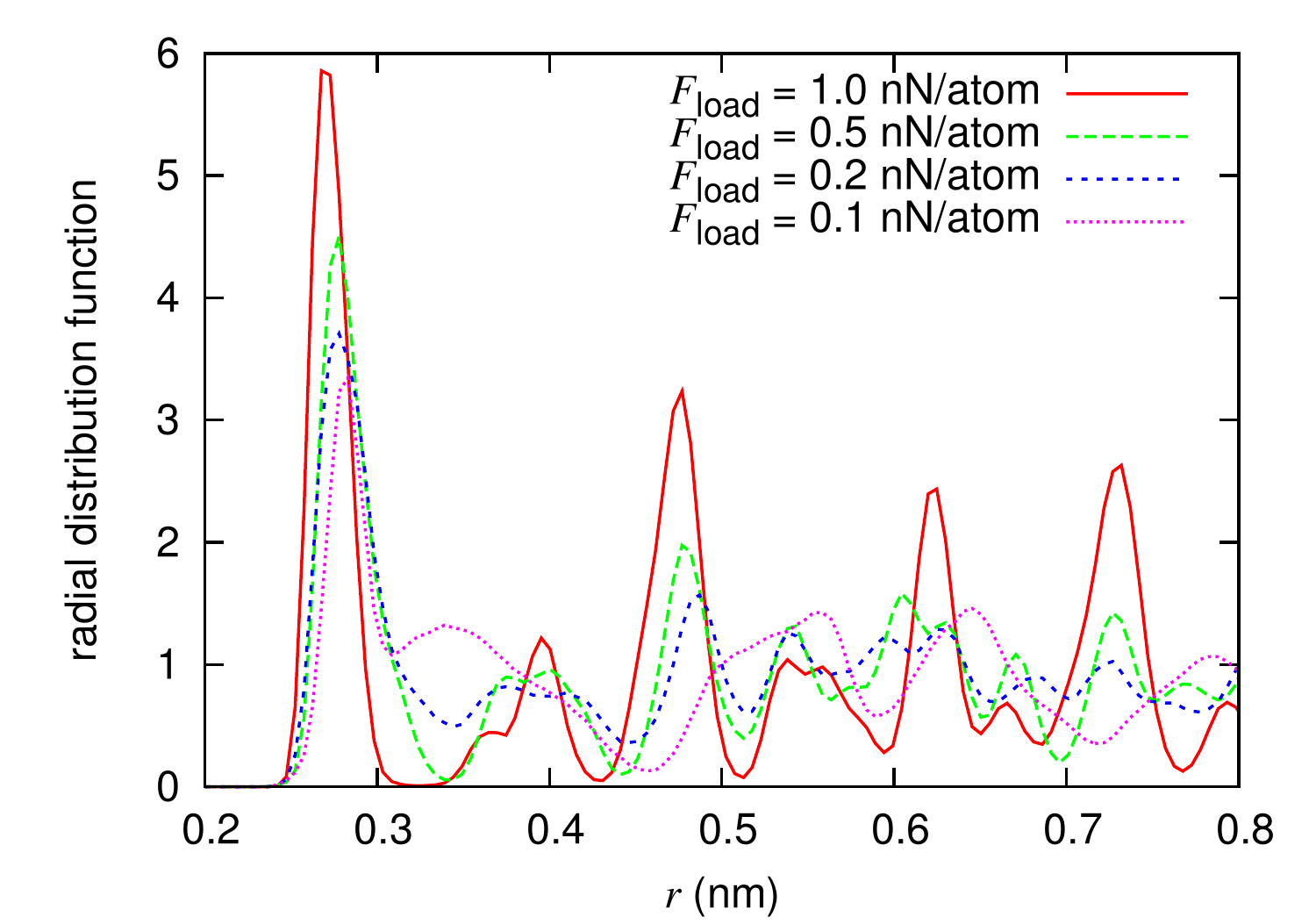,angle=270,width=8.6cm}\\
~~~(b)\hfill\strut\\[-5ex]
\epsfig{figure=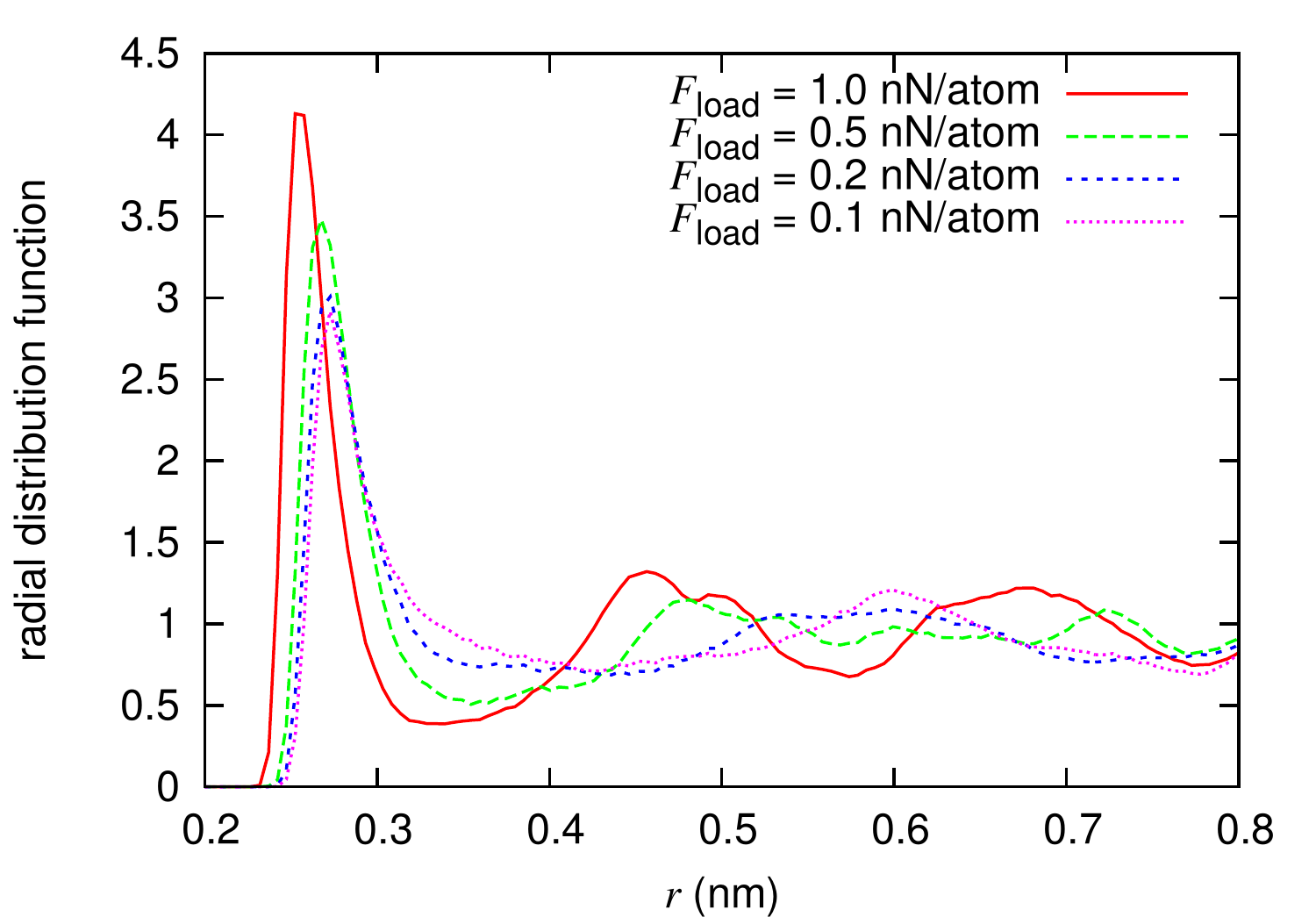,angle=270,width=8.6cm}
\caption{Oxygen-oxygen radial distribution functions obtained from the temperature-scan method (a) and from the load-scan method (b). The temperature-scan method results in more ice-like structures with rather well-defined peaks in the RDF while the load-scan method results in significantly more disordered, liquid-like environments.
In both these simulations $N_\mathrm{O}/N_\mathrm{C}=1.33$.
\label{fig:rdf}
}
\end{figure}

In order to better understand the different structures, we investigate the radial distribution function (RDF) and several other order parameters for a number of conditions.
The RDFs are plotted in Fig.~\ref{fig:rdf}.
Because of the confining geometry, the RDF is normalised not by $4 \pi r^2 N/(L_x L_y L_z)$, but by $ 4 \pi r^2 f(r,d) N/(L_x L_y d)$ with $f(r,d) = d/2r$ if $r>d$ and $f(r,d) = (1-r/2d)$ if $r<d$.
This is the RDF one would find for a slab of density $N/(L_x L_y d)$ and height $d$ in the $z$ direction.
The height $d$ was estimated from the maximum and minimum $z$ coordinates of all oxygen atoms.

For the system that results from the temperature-scan approach, the RDFs at higher loads show the typical isolated peaks of a crystalline structure. Interestingly, at the lowest load of 0.1 nN/atom the RDF is very similar to that found for the first layer of liquid TIP4P/2005 water on a BaF$_2$ substrate \cite{Kaya-BaF2-2013} with significant intensity at the interstitial distance around 0.35~nm and a near-complete loss of the second coordination shell around 0.45~nm. In ref.~\cite{Kaya-BaF2-2013} this was interpreted as a high-density form which was fully consistent with the experimental x-ray absorption spectra.
A difference between those results and ours is the presence here of significantly more structured peaks at intermediate distances.

Turning to the simulations with ramped-up load at constant temperature, we find much more smeared-out RDFs with structure consistent with a high-density liquid~\cite{Soper-HDL-LDL-2000}; only at the highest load do we find well-developed shell-structure, but still with peaks that are significantly broader than for the temperature scan simulations. Again, we find strong similarities between the present results and those of ref.~\cite{Kaya-BaF2-2013} for the high-density liquid and NaCl solution. In both the latter cases there is enhanced intensity in the interstitial region, a near-complete loss of the second shell and a build-up of intensity around 0.6~nm. 

\begin{figure}
\vskip1.0\bigskipamount
(a)~~~~~~~~~~~~~~~~~~~~~~~~~~~~~~~~~~~~\\[-2.5\bigskipamount]
\epsfig{figure=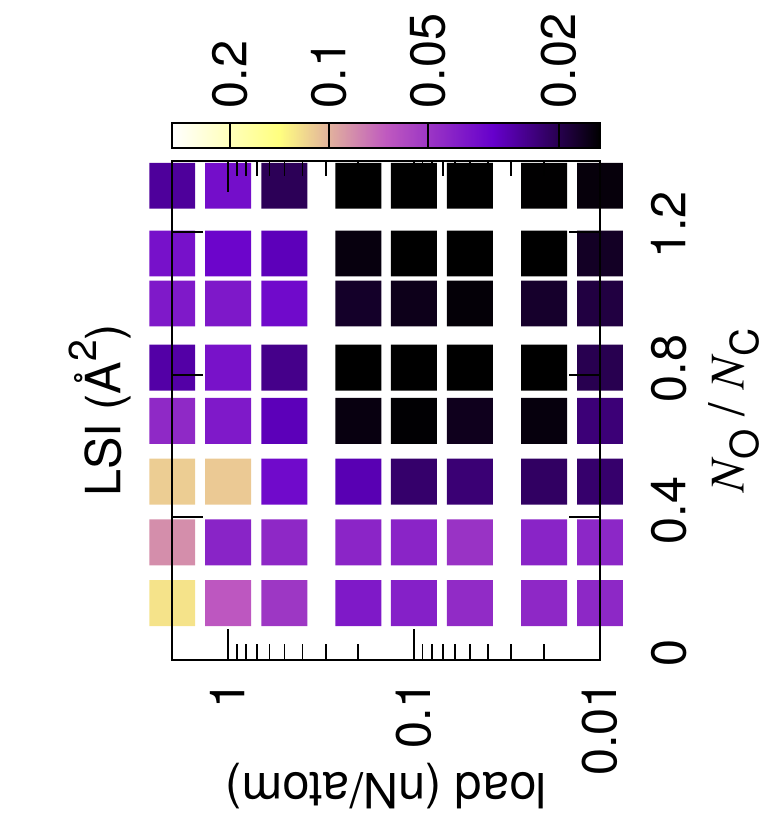,angle=270,width=4.25cm}\\
\vskip1.0\bigskipamount
~~(b)~~~~~~~~~~~~~~~~~~~~~~~~~~~~~~~~~~~~~~~(c)~~~~~~~~~~~~~~~~~~~~~~~~~~~~~~~~~~~~~~~
\\[-2.5\bigskipamount]
\epsfig{figure=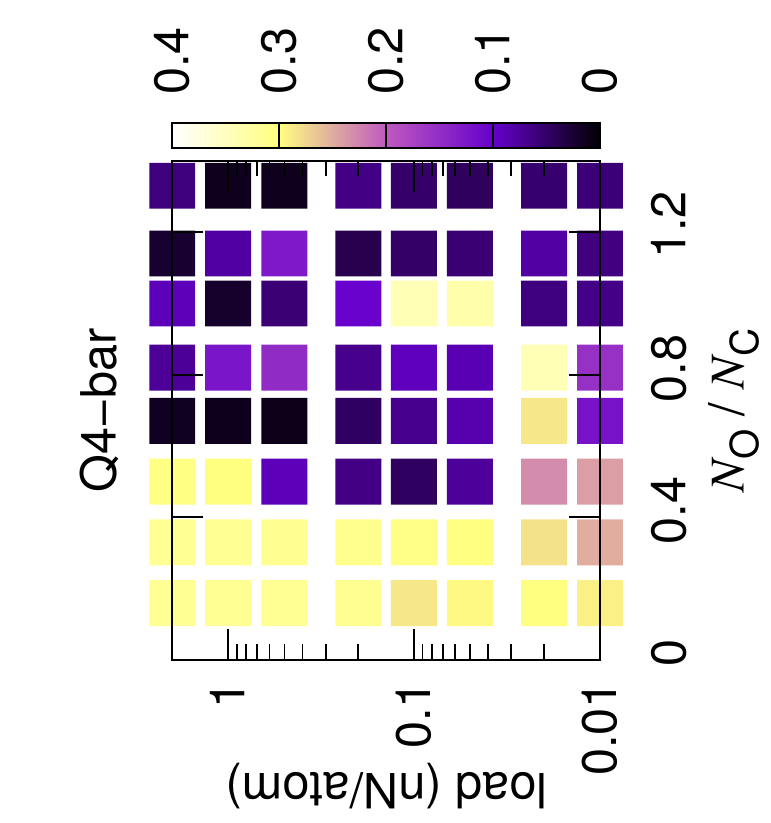,angle=270,width=4.25cm}
\epsfig{figure=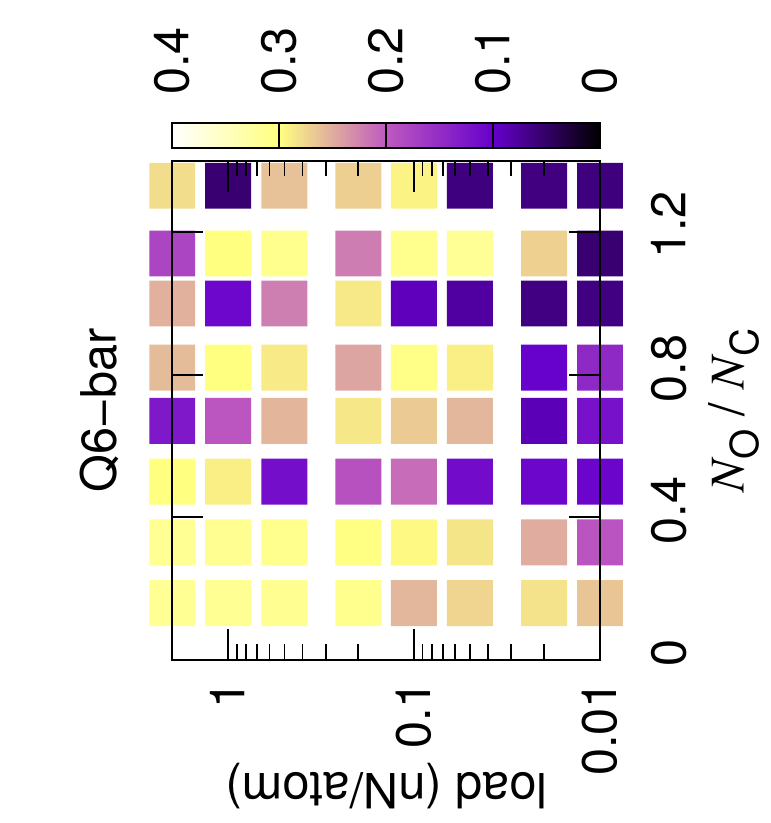,angle=270,width=4.25cm}
\caption{
Thickness-pressure plot of the average Local Structure Index (LSI), $\bar{q}4$, and $\bar{q}6$.
For comparison, 1~nN/atom of load corresponds to a pressure of 38~GPa.
\label{fig:phasediagram2}
}
\end{figure}

In Fig.~\ref{fig:phasediagram2} we classify the structure(s) in more detail through three order parameters as a function of the number of water molecules and the load, the  Local Structure Index (LSI) \cite{Shiratani1996,Shiratani1998} and the local bond order parameters $\bar{q}4$ and $\bar{q}6$~\cite{dellagobondorderparams}.

The LSI measures the degree of tetrahedral order versus disorder around a water molecule out to the second shell where the cutoff is set at 0.37~nm~\cite{Shiratani1996,Shiratani1998}. It is defined as $I(i) = {1\over n(i)}\sum_{j=1}^{n(i)}[\Delta(j;i)-\bar\Delta(i)]^2$ where $n(i)$ is the number of oxygen atoms out to the cutoff distance. These are ordered according to the distance from the central oxygen as $r_{i1}<r_{i2}<...< r_{in(i)}, \Delta(j;i) = r_{i,j+1}-r_{i,j}$ is the radial distance between the ordered oxygen neighbors and $\bar\Delta(i)$ is the mean of the sequential distances around the oxygen in molecule $i$. For a very structured tetrahedral environment there will be a large first distance followed by three very small distances and then a jump to the second shell. This situation will give a large squared deviation from the mean and thus a large LSI value while a more disordered local structure will give a low value.

We determine the symmetry of the phases using the local bond order parameters $\bar{q}4$ and $\bar{q}6$~\cite{dellagobondorderparams} applied to the oxygen atoms.
These are based on the Steinhardt bond-order parameters~\cite{steinhardtbondorderparams}, but are more sensitive to the local structure.
They can be written as a local sum over spherical harmonics $Y_{lm}$ of the relative positions between the particles $\vec{r}_{ij}$ as
\begin{align}
\bar{q}_l(i) = \sqrt{\frac{4 \pi}{2 l +1}\sum_{m=-l}^{m=l} | \bar{q}_{lm}(i) |^2}~,\\
\bar{q}_{lm}(u) = \frac{1}{N_b(i)+1} \sum_{k \in \mathrm{nn} \cup ~i} q_{lm}(k)~,\\
q_{lm}(i) = \frac{1}{N_b(i)} \sum_{j\in \mathrm{nn}} Y_{lm}(\vec{r}_{ij})~,
\end{align}
where $\mathrm{nn}$ indicates the set of nearest neighbors of particle $i$ and $N_b(i)$ is the number of nearest neighbors.
Nearest neighbors are defined as particles within a specific distance from each other, in our case 0.37~nm.
The quantities $\bar{q}_4(i)$ and $\bar{q}_6(i)$ are averaged over all oxygen atoms to obtain $\bar{q}_4$ and $\bar{q}_6$ respectively.

At low coverage, both $\bar{q}_4$ and $\bar{q}_6$ are high, as is the LSI.
This indicates that there is crystalline structure in the system, but it is not consistent with a specific bulk symmetry.  This is related to the fact that at low coverages there are only one or two layers of water molecules and the order parameters $\bar{q}_4$ and $\bar{q}_6$ are intended for use in bulk materials.
The LSI plot shows low values for loads up to 0.2~nN/atom and coverage $N_O/N_C$ above 0.6.
This indicates strongly disordered local environments for this range of parameters. As expected, increasing the load leads to somewhat higher LSI values, {\it i.e.} slightly more order as the interaction with the graphene sheets is strengthed, with the highest values found for the highest loads at the lowest coverages.
Closer visual inspection shows that the oxygen atoms in this regime are arranged in a single layer with inlayer structure.
The distance between the oxygen atoms is approximately 0.27~nm and the atoms are arranged in a square grid.
As the pressure increases, the structure adapts more to the graphite substrate, which distorts it from square to diamond-like.
 
At high loads and high coverage, the low LSI values indicate disorder, while the combination of high $\bar{q}_6$ and low $\bar{q}_4$ indicates an underlying BCC symmetry~\cite{dellagobondorderparams}.
We have confirmed this by visual inspection.
The distance between the carbon atoms within one layer is approximately 0.39~nm.  
The single-layer structure with spacing 0.27~nm is likely the (0,1,1) face of this BCC structure.

\section{Discussion}

Our results can be interpreted by comparison to the phase diagram of bulk ice, which is very well described by the TIP4P/2005 force-field which we use in the present study~\cite{TIP4P-2005,aragones2009}.
The crystalline structure we observe for the temperature-scan method is consistent with bulk ice VII as expected from the phase diagram of TIP4P/2005.
At room temperature, water can freeze at pressures above 0.9~GPa, forming tetragonal ice VI.  Above about 2~GPa the stable phase is ice VII, which has a BCC arrangement of oxygen atoms~\cite{iceVII}.  Above about 60~GPa, ice VII undergoes a continuous transition to ice X that is related to a rearrangement of the hydrogens.
It is interesting to note that ice VII must be formed by reducing the temperature of water at high pressures, rather than by increasing the pressure at ambient temperature.

Our simulations go up to significantly higher pressures than those of~\cite{watergraphenesims,watergraphenesimsstructurelessplates,watergraphenesimsDFT,squareicegeimnature,chenstickslipcontrol}, all of which consider otherwise similar confining geometries.
While the pressures we study in this work may seem high, they are in fact realistic, both in macroscopic contacts and in AFM experiments.
Due to the surface roughness of macroscopic contacting surfaces, the actual area of contact is small.
The contact pressures can be estimated using Amontons' laws and depend mostly on the material hardness.
Steel-on-steel contacts, for example, have contact pressures of about 1-3~GPa, regardless of the applied load, sufficient for the formation of ice VII.
While we do not know precisely the size of the AFM tip in the experiments by Jinesh et al.~\cite{jinesh2006,jinesh2008}, the length scale of the tip roughness should be around a nm.
Combined with the experimental loads of several nN this leads to local pressure peaks in the range of several GPa, and thus also sufficient for the formation of ice VII.

The previous simulations did not vary the temperature during equilibration.
This is likely why the confined ice VII structure that we find was not observed.
The authors of Ref.~\cite{watergraphenesims} already noted that at higher pressures they were not able to obtain convergence, consistent with our observations of the difference between the load and temperature scans.
The transmission electron microscopy (TEM) experiments of Ref.~\cite{squareicegeimnature} were accompanied by simulations at pressures up to 1~GPa.
Contrary to the TEM experiments, those simulations indicate an FCC-like lattice.
Given the setup of the simulations, we believe that this FCC structure may have been related either to ice VI, or to the rhombic structure found at extremely low pressure in other simulations~\cite{watergraphenesims}.
While the authors of Ref.~\cite{watergraphenesims} did not evaluate this possibility explicitely, it appears that their TEM measurements would be consistent with square ice with a multi-layer BCC lattice.

\begin{figure}
\epsfig{figure=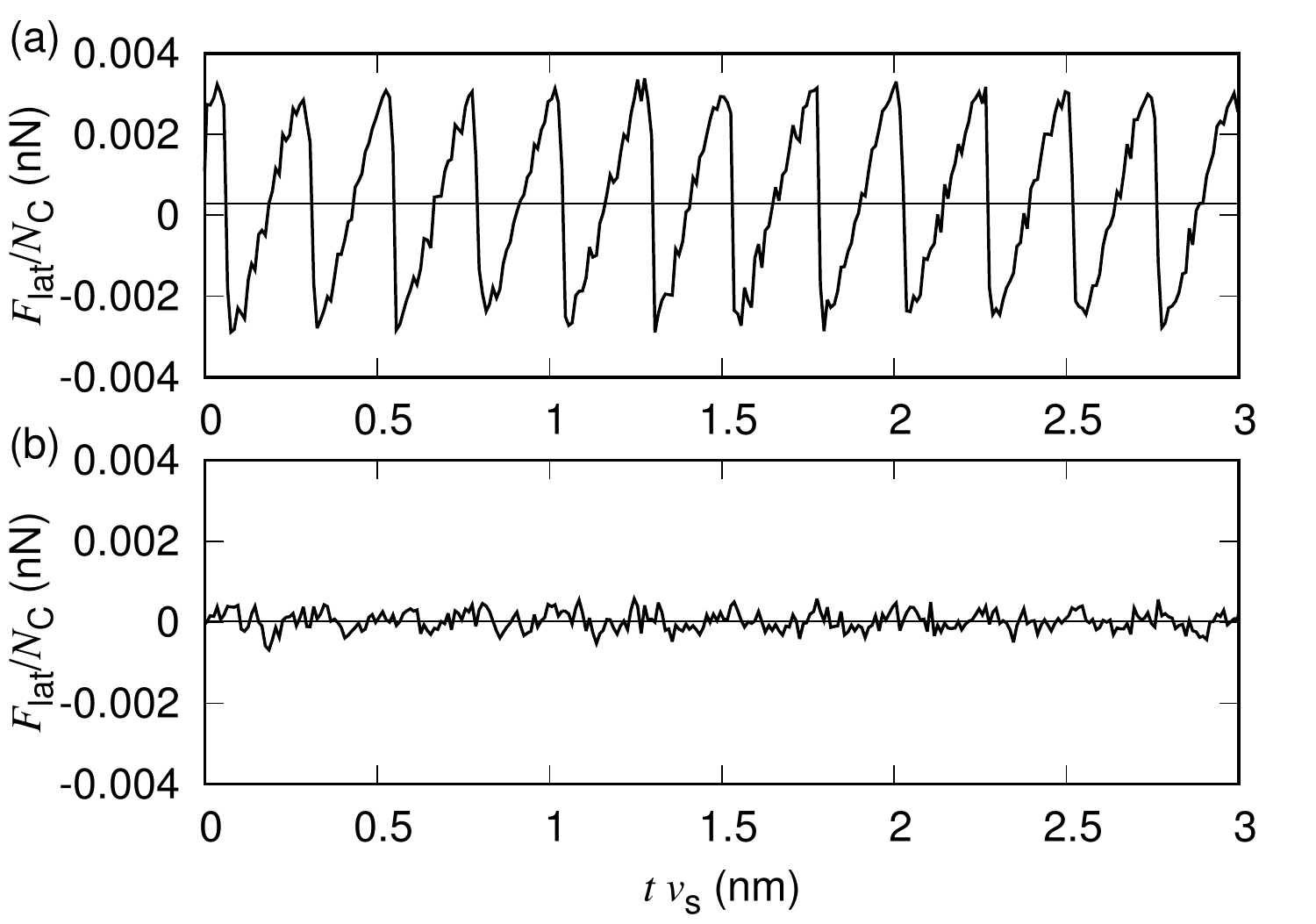, angle=270,width=8.6cm}
\caption{
Force traces for identical parameters (boundary layer thickness, sliding velocity, etc) but different initial conditions, liquid-like (a) and ice VII-like (b).
Due to the incommensurate lattice parameters of ice VII and graphite, there is in this case superlubric sliding.  The liquid-like water rearranges in a configuration that is strongly commensurate with the graphite, and thus has higher friction.  The stick-slip period is consistent with the graphite substrate.
In these simulations $N_\mathrm{O}/N_\mathrm{C}=1.33$ and the load was (ramped up to) 0.5~nN/atom.
The average lateral forces were $0.29\pm0.02$~pN/atom for the liquid-like water and $0.027\pm0.007$~pN/atom for the ice VII-like water.  The errors were estimated using block averaging.
The plotted forces were averaged over short intervals of 5~ps to eliminate thermal fluctuations.
\label{fig:forcetraces}
}
\end{figure}

We also briefly discuss squeezing out of the water.
To investigate this we have performed simulations of a larger system, where the bottom sheets were 1.5 times larger in both directions, while the top sheets were kept at the same size.
The water could thus escape laterally from the contact asperity through the open sides.
We then slowly moved the plates together and recorded the force needed to squeeze out the simulated water molecules into the vacuum above the top plate.
The load needed to push out the last layer corresponds to about 0.1~nN/atom, which equates to approximately 4~GPa, too low for the formation of bulk ice VII.
The extreme nonequilibrium nature of friction, and the long time scales associated with squeezing out the liquid, can lead to much higher pressures in real contacts.

\section{Friction and comparison to AFM experiments}

When crystalline solids come into contact there exists the possibility of extremely low friction if the lattice parameters are incommensurate (mismatched)~\cite{vanishingstaticfriction}.
This phenomenon is known as structural superlubricity and has been widely studied.
Specifically, it has been shown both theoretically~\cite{mueserprl,astridgoldgraphite} and experimentally~\cite{dietzelscaling} that incommensurate crystal-on-crystal contacts generally have much lower friction than amorphous-on-crystalline contacts.

Ice VII and graphene constitute such an incommensurate crystalline combination with mismatched lattice parameters and we have checked that this indeed leads to low friction.
In Fig.~\ref{fig:forcetraces} we show the lateral force as a function of time for two samples, one with liquid-like order, and one with ice VII-like, when sliding at a relative speed of 2~m/s.
Due to the low-friction slide plane between the ice and graphene, the friction for the ice is lower.
In the liquid-like system, there is a much stronger signature of stick-slip, with a period that is consistent with the lattice period of graphene.

In the experiments~\cite{jinesh2008} at high velocities the same stick-slip was observed that we see in our simulations of sliding with liquid-like confined water, i.\ e.\ with the lattice period of graphite.
At high velocities we expect that the rapid increase in the pressure under the tip is best represented by the load-scan simulations, and thus we indeed expect liquid-like structure under those conditions.

At low velocities, the experiments~\cite{jinesh2008}  scanning a sharp tungsten tip over a wet graphite surface,  show a high friction with a stick-slip period different from the lattice period of graphite.
This lattice period (0.38$\pm$0.03~nm) is consistent with the period of 0.35~nm that we find for the crystalline ice structure that results from the temperature-scan equilibration method.
We thus conclude that very probably ice VII is formed in these experiments.

In these experiments, however, its presence in the contact does not lead to superlubric sliding, but rather to high friction.
This may be due to the difference in configuration between the experiments and our simulations.
In the experiments, the possible slide planes are severely restricted.
The graphite substrate in the experiments is probably (nearly) atomically flat over the distance that the tip travels, and the ice could, in principle, slide easily with respect to the surface.
This could however be prevented because in order for the block of ice under the tip to slide, it would need to displace the liquid water surrounding it.
The surface roughness of the tip also prevents the tip from sliding easily with respect to the ice.
We suspect that as a result of this, in the experiments there is no clear slide plane as there is in the somewhat idealised solutions.
Instead, the ice has to be fractured, and ice slides against ice with the same lattice constants, which produces high friction and a stick-slip period corresponding to ice, not graphite.
When graphite is used as a solid lubricant, however, water is sandwhiched between flakes of (nearly) atomically flat graphite or graphene.
With regard to which slide planes are possible, the situation in practice with humid graphite as lubricant may be more similar to our simulations than to the experiments using an AFM tip.

\section{Conclusions}
The phase diagram of bulk water is already complicated, but in the present case, it appears that the extreme confinement of a few layers actually does not increase the level of complexity drastically.
We find from our simulations that the experiments for water on graphite~\cite{jinesh2006,jinesh2008} and graphene~\cite{squareicegeimnature} can be explained by considering the formation of ice VII.
This ice phase is formed at extremely high pressures, but it is metastable under a wide range of conditions, including ambient~\cite{metastableiceVII}.
In our simulations, its formation depends strongly on how the system is treated, which is consistent with the velocity-dependence found in experiments~\cite{jinesh2008}.

We suspect that it is mainly the high pressure exerted by the graphene sheets in the simulation that gives the stability of ice VII: the structure is incommensurate with the graphene and under the simulated conditions ice VII is the stable phase in the phase diagram of water.
Moreover, the similarities between our simulation results and measurements performed on water on a BaF$_2$ substrate~\cite{Kaya-BaF2-2013}, as well as on graphene on Pt(111)~\cite{kimmel2009}, further support an interpretation based on the properties of water rather than those of the surface. We note further that the water-water interaction is significantly stronger than the water-graphene interaction, leading to the hydrogens preferentially interacting with water rather than the graphene~\cite{kimmel2009}

Due to the difference in lattice parameters between the ice VII and graphite, the two slide easily with respect to each other.
This means that the formation of ice VII between graphene layers has the potential to drastically lower friction through the mechanism of structural superlubricity.
We have confirmed this in our simulations.
For structural superlubricity to happen requires atomically nearly flat surfaces.
In the AFM experiments~\cite{jinesh2006,jinesh2008}, this sliding was prevented by the geometry of the contact.
Nevertheless, in a solid lubricant, nearly atomically flat graphite and graphene flakes slide with respect to one-another, and it may be possible for square ice VII to form, resulting in graphene  sliding on an incommensurate crystalline ice layer, thus reducing the friction.
Moreover, the pressures needed for the formation of ice VII are common in microscopic contact asperities under realistic conditions.

We therefore conclude that the formation of ice VII between layers of graphene or graphite is a possible explanation as to why graphite is much more effective as a solid lubricant in humid conditions than in dry conditions.

\begin{acknowledgments}
ASdW acknowledges support from the Swedish Research Council (Vetenskapsr\aa{}det), the COST action MP1303, and the Swedish National Infrastructure for Computing (SNIC).
\end{acknowledgments}


\begin{thebibliography}{41}
\expandafter\ifx\csname natexlab\endcsname\relax\def\natexlab#1{#1}\fi
\expandafter\ifx\csname bibnamefont\endcsname\relax
  \def\bibnamefont#1{#1}\fi
\expandafter\ifx\csname bibfnamefont\endcsname\relax
  \def\bibfnamefont#1{#1}\fi
\expandafter\ifx\csname citenamefont\endcsname\relax
  \def\citenamefont#1{#1}\fi
\expandafter\ifx\csname url\endcsname\relax
  \def\url#1{\texttt{#1}}\fi
\expandafter\ifx\csname urlprefix\endcsname\relax\def\urlprefix{URL }\fi
\providecommand{\bibinfo}[2]{#2}
\providecommand{\eprint}[2][]{\url{#2}}

\bibitem[{\citenamefont{Donnet and
  Erdemir}(2004)}]{solidlubricantcoatingsreview}
\bibinfo{author}{\bibfnamefont{C.}~\bibnamefont{Donnet}} \bibnamefont{and}
  \bibinfo{author}{\bibfnamefont{A.}~\bibnamefont{Erdemir}},
  \bibinfo{journal}{Tribol.~Lett.} \textbf{\bibinfo{volume}{17}},
  \bibinfo{pages}{389} (\bibinfo{year}{2004}).

\bibitem[{\citenamefont{de~Wijn et~al.}(2011)\citenamefont{de~Wijn, Fasolino,
  Filippov, and Urbakh}}]{onssolidlubricant}
\bibinfo{author}{\bibfnamefont{A.~S.} \bibnamefont{de~Wijn}},
  \bibinfo{author}{\bibfnamefont{A.}~\bibnamefont{Fasolino}},
  \bibinfo{author}{\bibfnamefont{A.~E.} \bibnamefont{Filippov}},
  \bibnamefont{and} \bibinfo{author}{\bibfnamefont{M.}~\bibnamefont{Urbakh}},
  \bibinfo{journal}{Europhysics Lett.} \textbf{\bibinfo{volume}{95}},
  \bibinfo{pages}{66002} (\bibinfo{year}{2011}).

\bibitem[{\citenamefont{Berman et~al.}(2015)\citenamefont{Berman, Deshmukh,
  Sankaranarayanan1, Erdemir, and Sumant}}]{graphenenanoscrolls}
\bibinfo{author}{\bibfnamefont{D.}~\bibnamefont{Berman}},
  \bibinfo{author}{\bibfnamefont{S.~A.} \bibnamefont{Deshmukh}},
  \bibinfo{author}{\bibfnamefont{S.~K.~R.~S.} \bibnamefont{Sankaranarayanan1}},
  \bibinfo{author}{\bibfnamefont{A.}~\bibnamefont{Erdemir}}, \bibnamefont{and}
  \bibinfo{author}{\bibfnamefont{A.~V.} \bibnamefont{Sumant}},
  \bibinfo{journal}{Science} \textbf{\bibinfo{volume}{348}},
  \bibinfo{pages}{1118} (\bibinfo{year}{2015}).

\bibitem[{\citenamefont{Cerveny et~al.}(2016)\citenamefont{Cerveny, Mallamace,
  Swenson, Vogel, and Xu}}]{ChemRev2016}
\bibinfo{author}{\bibfnamefont{S.}~\bibnamefont{Cerveny}},
  \bibinfo{author}{\bibfnamefont{F.}~\bibnamefont{Mallamace}},
  \bibinfo{author}{\bibfnamefont{J.}~\bibnamefont{Swenson}},
  \bibinfo{author}{\bibfnamefont{M.}~\bibnamefont{Vogel}}, \bibnamefont{and}
  \bibinfo{author}{\bibfnamefont{L.}~\bibnamefont{Xu}}, \bibinfo{journal}{Chem.
  Rev.} \textbf{\bibinfo{volume}{116}}, \bibinfo{pages}{7608}
  (\bibinfo{year}{2016}), \urlprefix\url{doi:10.1021/acs.chemrev.5b00609}.

\bibitem[{\citenamefont{Algara-Siller et~al.}(2015)\citenamefont{Algara-Siller,
  Lehtinen, Wang, Nair, Kaiser, Wu, Geim, and
  Grigorieva}}]{squareicegeimnature}
\bibinfo{author}{\bibfnamefont{G.}~\bibnamefont{Algara-Siller}},
  \bibinfo{author}{\bibfnamefont{O.}~\bibnamefont{Lehtinen}},
  \bibinfo{author}{\bibfnamefont{F.~C.} \bibnamefont{Wang}},
  \bibinfo{author}{\bibfnamefont{R.~R.} \bibnamefont{Nair}},
  \bibinfo{author}{\bibfnamefont{U.}~\bibnamefont{Kaiser}},
  \bibinfo{author}{\bibfnamefont{H.~A.} \bibnamefont{Wu}},
  \bibinfo{author}{\bibfnamefont{A.~K.} \bibnamefont{Geim}}, \bibnamefont{and}
  \bibinfo{author}{\bibfnamefont{I.~V.} \bibnamefont{Grigorieva}},
  \bibinfo{journal}{Nature} \textbf{\bibinfo{volume}{519}},
  \bibinfo{pages}{443} (\bibinfo{year}{2015}).

\bibitem[{\citenamefont{Xu et~al.}(2010)\citenamefont{Xu, Cao, and
  Heath}}]{grapheneonmica}
\bibinfo{author}{\bibfnamefont{K.}~\bibnamefont{Xu}},
  \bibinfo{author}{\bibfnamefont{P.}~\bibnamefont{Cao}}, \bibnamefont{and}
  \bibinfo{author}{\bibfnamefont{J.~R.} \bibnamefont{Heath}},
  \bibinfo{journal}{Science} \textbf{\bibinfo{volume}{329}},
  \bibinfo{pages}{1188} (\bibinfo{year}{2010}).

\bibitem[{\citenamefont{Severin et~al.}(2012)\citenamefont{Severin, Lange,
  Sokolov, and Rabe}}]{Severin2012}
\bibinfo{author}{\bibfnamefont{N.}~\bibnamefont{Severin}},
  \bibinfo{author}{\bibfnamefont{P.}~\bibnamefont{Lange}},
  \bibinfo{author}{\bibfnamefont{I.~M.} \bibnamefont{Sokolov}},
  \bibnamefont{and} \bibinfo{author}{\bibfnamefont{J.~P.} \bibnamefont{Rabe}},
  \bibinfo{journal}{Nano Lett.} \textbf{\bibinfo{volume}{12}},
  \bibinfo{pages}{774} (\bibinfo{year}{2012}).

\bibitem[{\citenamefont{Verdaguer et~al.}(2013)\citenamefont{Verdaguer, Segura,
  L\'opez-Mir, Sauthier, and Fraxedas}}]{verdaguer2013}
\bibinfo{author}{\bibfnamefont{A.}~\bibnamefont{Verdaguer}},
  \bibinfo{author}{\bibfnamefont{J.~J.} \bibnamefont{Segura}},
  \bibinfo{author}{\bibfnamefont{L.}~\bibnamefont{L\'opez-Mir}},
  \bibinfo{author}{\bibfnamefont{G.}~\bibnamefont{Sauthier}}, \bibnamefont{and}
  \bibinfo{author}{\bibfnamefont{J.}~\bibnamefont{Fraxedas}},
  \bibinfo{journal}{J. Chem. Phys.} \textbf{\bibinfo{volume}{138}},
  \bibinfo{pages}{121101} (\bibinfo{year}{2013}).

\bibitem[{\citenamefont{Kimmel et~al.}(2009)\citenamefont{Kimmel, Matthiesen,
  Baer, Mundy, Petrik, Dohn{\'a}lek, and Kay}}]{kimmel2009}
\bibinfo{author}{\bibfnamefont{G.~A.} \bibnamefont{Kimmel}},
  \bibinfo{author}{\bibfnamefont{J.}~\bibnamefont{Matthiesen}},
  \bibinfo{author}{\bibfnamefont{M.}~\bibnamefont{Baer}},
  \bibinfo{author}{\bibfnamefont{J.}~\bibnamefont{Mundy},
  \bibfnamefont{Christopher}}, \bibinfo{author}{\bibfnamefont{N.~G.}
  \bibnamefont{Petrik}},
  \bibinfo{author}{\bibfnamefont{Z.}~\bibnamefont{Dohn{\'a}lek}},
  \bibnamefont{and} \bibinfo{author}{\bibfnamefont{B.~D.} \bibnamefont{Kay}},
  \bibinfo{journal}{J. Am. Chem. Soc.} \textbf{\bibinfo{volume}{131}},
  \bibinfo{pages}{12838} (\bibinfo{year}{2009}).

\bibitem[{\citenamefont{Vilhena et~al.}(2016)\citenamefont{Vilhena, Pimentel,
  Pedraz, Luo, Serena, Pina, Gnecco, and P{\'e}rez}}]{vilhena2016}
\bibinfo{author}{\bibfnamefont{J.~G.} \bibnamefont{Vilhena}},
  \bibinfo{author}{\bibfnamefont{C.}~\bibnamefont{Pimentel}},
  \bibinfo{author}{\bibfnamefont{P.}~\bibnamefont{Pedraz}},
  \bibinfo{author}{\bibfnamefont{F.}~\bibnamefont{Luo}},
  \bibinfo{author}{\bibfnamefont{P.~A.} \bibnamefont{Serena}},
  \bibinfo{author}{\bibfnamefont{C.~M.} \bibnamefont{Pina}},
  \bibinfo{author}{\bibfnamefont{E.}~\bibnamefont{Gnecco}}, \bibnamefont{and}
  \bibinfo{author}{\bibfnamefont{R.}~\bibnamefont{P{\'e}rez}},
  \bibinfo{journal}{ACS Nano} \textbf{\bibinfo{volume}{10}},
  \bibinfo{pages}{4288} (\bibinfo{year}{2016}).

\bibitem[{\citenamefont{Chen et~al.}(2016)\citenamefont{Chen, Schusteritsch,
  Pickard, Salzmann, and Michaelides}}]{chen2diceprl}
\bibinfo{author}{\bibfnamefont{J.}~\bibnamefont{Chen}},
  \bibinfo{author}{\bibfnamefont{G.}~\bibnamefont{Schusteritsch}},
  \bibinfo{author}{\bibfnamefont{C.~J.} \bibnamefont{Pickard}},
  \bibinfo{author}{\bibfnamefont{C.~G.} \bibnamefont{Salzmann}},
  \bibnamefont{and}
  \bibinfo{author}{\bibfnamefont{A.}~\bibnamefont{Michaelides}},
  \bibinfo{journal}{Phys. Rev. Lett.} \textbf{\bibinfo{volume}{116}},
  \bibinfo{pages}{025501} (\bibinfo{year}{2016}).

\bibitem[{\citenamefont{Mosaddeghi et~al.}(2012)\citenamefont{Mosaddeghi,
  Alavi, Kowsari, and Najafi}}]{watergraphenesims}
\bibinfo{author}{\bibfnamefont{H.}~\bibnamefont{Mosaddeghi}},
  \bibinfo{author}{\bibfnamefont{S.}~\bibnamefont{Alavi}},
  \bibinfo{author}{\bibfnamefont{M.~H.} \bibnamefont{Kowsari}},
  \bibnamefont{and} \bibinfo{author}{\bibfnamefont{B.}~\bibnamefont{Najafi}},
  \bibinfo{journal}{J. Chem. Phys.} \textbf{\bibinfo{volume}{137}},
  \bibinfo{eid}{184703} (\bibinfo{year}{2012}),
  \urlprefix\url{http://scitation.aip.org/content/aip/journal/jcp/137/18/10.1063/1.4763984}.

\bibitem[{\citenamefont{Cicero et~al.}(2008)\citenamefont{Cicero, Grossman,
  Schwegler, Gygi, and Galli}}]{watergraphenesimsDFT}
\bibinfo{author}{\bibfnamefont{G.}~\bibnamefont{Cicero}},
  \bibinfo{author}{\bibfnamefont{J.~C.} \bibnamefont{Grossman}},
  \bibinfo{author}{\bibfnamefont{E.}~\bibnamefont{Schwegler}},
  \bibinfo{author}{\bibfnamefont{F.}~\bibnamefont{Gygi}}, \bibnamefont{and}
  \bibinfo{author}{\bibfnamefont{G.}~\bibnamefont{Galli}}, \bibinfo{journal}{J.
  Am. Chem. Soc.} \textbf{\bibinfo{volume}{130}}, \bibinfo{pages}{1871}
  (\bibinfo{year}{2008}), \bibinfo{note}{pMID: 18211065},
  \eprint{http://dx.doi.org/10.1021/ja074418+},
  \urlprefix\url{http://dx.doi.org/10.1021/ja074418+}.

\bibitem[{\citenamefont{Han et~al.}(2010)\citenamefont{Han, Choi, Kumar, and
  Stanley}}]{watergraphenesimsstructurelessplates}
\bibinfo{author}{\bibfnamefont{S.}~\bibnamefont{Han}},
  \bibinfo{author}{\bibfnamefont{M.~Y.} \bibnamefont{Choi}},
  \bibinfo{author}{\bibfnamefont{P.}~\bibnamefont{Kumar}}, \bibnamefont{and}
  \bibinfo{author}{\bibfnamefont{H.~E.} \bibnamefont{Stanley}},
  \bibinfo{journal}{Nature Physics} \textbf{\bibinfo{volume}{6}},
  \bibinfo{pages}{685} (\bibinfo{year}{2010}).

\bibitem[{\citenamefont{Giovambattista
  et~al.}(2009)\citenamefont{Giovambattista, Rossky, and
  Debenedetti}}]{giovambattista2009}
\bibinfo{author}{\bibfnamefont{N.}~\bibnamefont{Giovambattista}},
  \bibinfo{author}{\bibfnamefont{P.~J.} \bibnamefont{Rossky}},
  \bibnamefont{and} \bibinfo{author}{\bibfnamefont{P.~G.}
  \bibnamefont{Debenedetti}}, \bibinfo{journal}{Phys. Rev. Lett.}
  \textbf{\bibinfo{volume}{102}}, \bibinfo{pages}{050603}
  (\bibinfo{year}{2009}).

\bibitem[{\citenamefont{Tocci et~al.}(2014)\citenamefont{Tocci, Joly, and
  Michaelides}}]{tocci2014}
\bibinfo{author}{\bibfnamefont{G.}~\bibnamefont{Tocci}},
  \bibinfo{author}{\bibfnamefont{L.}~\bibnamefont{Joly}}, \bibnamefont{and}
  \bibinfo{author}{\bibfnamefont{A.}~\bibnamefont{Michaelides}},
  \bibinfo{journal}{Nano Lett.} \textbf{\bibinfo{volume}{14}},
  \bibinfo{pages}{6872} (\bibinfo{year}{2014}).

\bibitem[{\citenamefont{Koga et~al.}(2001)\citenamefont{Koga, Gao, Tanaka, and
  Zeng}}]{nanotubesquareice}
\bibinfo{author}{\bibfnamefont{K.}~\bibnamefont{Koga}},
  \bibinfo{author}{\bibfnamefont{G.~T.} \bibnamefont{Gao}},
  \bibinfo{author}{\bibfnamefont{H.}~\bibnamefont{Tanaka}}, \bibnamefont{and}
  \bibinfo{author}{\bibfnamefont{X.~C.} \bibnamefont{Zeng}},
  \bibinfo{journal}{Nature} \textbf{\bibinfo{volume}{412}},
  \bibinfo{pages}{802} (\bibinfo{year}{2001}).

\bibitem[{\citenamefont{Jinesh and Frenken}(2006)}]{jinesh2006}
\bibinfo{author}{\bibfnamefont{K.~B.} \bibnamefont{Jinesh}} \bibnamefont{and}
  \bibinfo{author}{\bibfnamefont{J.~W.~M.} \bibnamefont{Frenken}},
  \bibinfo{journal}{Phys. Rev. Lett.} \textbf{\bibinfo{volume}{96}},
  \bibinfo{pages}{166103} (\bibinfo{year}{2006}),
  \urlprefix\url{http://link.aps.org/doi/10.1103/PhysRevLett.96.166103}.

\bibitem[{\citenamefont{Chen et~al.}(2015)\citenamefont{Chen, Foster, Alava,
  and Laurson}}]{chenstickslipcontrol}
\bibinfo{author}{\bibfnamefont{W.}~\bibnamefont{Chen}},
  \bibinfo{author}{\bibfnamefont{A.~S.} \bibnamefont{Foster}},
  \bibinfo{author}{\bibfnamefont{M.~J.} \bibnamefont{Alava}}, \bibnamefont{and}
  \bibinfo{author}{\bibfnamefont{L.}~\bibnamefont{Laurson}},
  \bibinfo{journal}{Phys. Rev. Lett.} \textbf{\bibinfo{volume}{114}},
  \bibinfo{pages}{095502} (\bibinfo{year}{2015}),
  \urlprefix\url{http://link.aps.org/doi/10.1103/PhysRevLett.114.095502}.

\bibitem[{\citenamefont{Dienwiebel et~al.}(2004)\citenamefont{Dienwiebel,
  Verhoeven, Pradeep, Frenken, Heimberg, and Zandbergen}}]{Dienwiebel2004}
\bibinfo{author}{\bibfnamefont{M.}~\bibnamefont{Dienwiebel}},
  \bibinfo{author}{\bibfnamefont{G.~S.} \bibnamefont{Verhoeven}},
  \bibinfo{author}{\bibfnamefont{N.}~\bibnamefont{Pradeep}},
  \bibinfo{author}{\bibfnamefont{J.~W.~M.} \bibnamefont{Frenken}},
  \bibinfo{author}{\bibfnamefont{J.~A.} \bibnamefont{Heimberg}},
  \bibnamefont{and} \bibinfo{author}{\bibfnamefont{H.~W.}
  \bibnamefont{Zandbergen}}, \bibinfo{journal}{Phys.~Rev.~Lett.}
  \textbf{\bibinfo{volume}{92}}, \bibinfo{pages}{126101}
  (\bibinfo{year}{2004}).

\bibitem[{\citenamefont{Jinesh and Frenken}(2008)}]{jinesh2008}
\bibinfo{author}{\bibfnamefont{K.~B.} \bibnamefont{Jinesh}} \bibnamefont{and}
  \bibinfo{author}{\bibfnamefont{J.~W.~M.} \bibnamefont{Frenken}},
  \bibinfo{journal}{Phys. Rev. Lett.} \textbf{\bibinfo{volume}{101}},
  \bibinfo{pages}{036101} (\bibinfo{year}{2008}),
  \urlprefix\url{http://link.aps.org/doi/10.1103/PhysRevLett.101.036101}.

\bibitem[{\citenamefont{Li et~al.}(2011)\citenamefont{Li, Dong, Perez, Martini,
  and Carpick}}]{lowspeedfrictionsimulations}
\bibinfo{author}{\bibfnamefont{Q.}~\bibnamefont{Li}},
  \bibinfo{author}{\bibfnamefont{Y.}~\bibnamefont{Dong}},
  \bibinfo{author}{\bibfnamefont{D.}~\bibnamefont{Perez}},
  \bibinfo{author}{\bibfnamefont{A.}~\bibnamefont{Martini}}, \bibnamefont{and}
  \bibinfo{author}{\bibfnamefont{R.~W.} \bibnamefont{Carpick}},
  \bibinfo{journal}{Phys. Rev. Lett.} \textbf{\bibinfo{volume}{106}},
  \bibinfo{pages}{126101} (\bibinfo{year}{2011}),
  \urlprefix\url{http://link.aps.org/doi/10.1103/PhysRevLett.106.126101}.

\bibitem[{\citenamefont{Plimpton}(1995)}]{lammps}
\bibinfo{author}{\bibfnamefont{S.~J.} \bibnamefont{Plimpton}},
  \bibinfo{journal}{J. Comp. Phys.} \textbf{\bibinfo{volume}{117}},
  \bibinfo{pages}{1} (\bibinfo{year}{1995}),
  \bibinfo{note}{http://lammps.sandia.gov}.

\bibitem[{\citenamefont{Abascal and Vega}(2005)}]{TIP4P-2005}
\bibinfo{author}{\bibfnamefont{J.~L.~F.} \bibnamefont{Abascal}}
  \bibnamefont{and} \bibinfo{author}{\bibfnamefont{C.}~\bibnamefont{Vega}},
  \bibinfo{journal}{J. Chem. Phys.} \textbf{\bibinfo{volume}{123}},
  \bibinfo{pages}{234505} (\bibinfo{year}{2005}).

\bibitem[{\citenamefont{Aragones et~al.}(2009)\citenamefont{Aragones, Conde,
  Noya, and Vega}}]{aragones2009}
\bibinfo{author}{\bibfnamefont{J.~L.} \bibnamefont{Aragones}},
  \bibinfo{author}{\bibfnamefont{M.~M.} \bibnamefont{Conde}},
  \bibinfo{author}{\bibfnamefont{E.~G.} \bibnamefont{Noya}}, \bibnamefont{and}
  \bibinfo{author}{\bibfnamefont{C.}~\bibnamefont{Vega}},
  \bibinfo{journal}{Phys. Chem. Phys. Chem.} \textbf{\bibinfo{volume}{11}},
  \bibinfo{pages}{543} (\bibinfo{year}{2009}).

\bibitem[{\citenamefont{Stuart et~al.}(2000)\citenamefont{Stuart, Tutein, and
  Harrison}}]{airebo}
\bibinfo{author}{\bibfnamefont{S.~J.} \bibnamefont{Stuart}},
  \bibinfo{author}{\bibfnamefont{A.~B.} \bibnamefont{Tutein}},
  \bibnamefont{and} \bibinfo{author}{\bibfnamefont{J.~A.}
  \bibnamefont{Harrison}}, \bibinfo{journal}{J. Chem. Phys.}
  \textbf{\bibinfo{volume}{112}}, \bibinfo{pages}{6472} (\bibinfo{year}{2000}),
  \urlprefix\url{http://scitation.aip.org/content/aip/journal/jcp/112/14/10.1063/1.481208}.

\bibitem[{\citenamefont{Reguzzoni et~al.}(2012)\citenamefont{Reguzzoni,
  Fasolino, Molinari, and Righi}}]{graphenecorrugationproblem}
\bibinfo{author}{\bibfnamefont{M.}~\bibnamefont{Reguzzoni}},
  \bibinfo{author}{\bibfnamefont{A.}~\bibnamefont{Fasolino}},
  \bibinfo{author}{\bibfnamefont{E.}~\bibnamefont{Molinari}}, \bibnamefont{and}
  \bibinfo{author}{\bibfnamefont{M.~C.} \bibnamefont{Righi}},
  \bibinfo{journal}{Phys. Rev. B} \textbf{\bibinfo{volume}{86}},
  \bibinfo{pages}{245434} (\bibinfo{year}{2012}),
  \urlprefix\url{http://link.aps.org/doi/10.1103/PhysRevB.86.245434}.

\bibitem[{\citenamefont{Persson}(2000)}]{perssonboek}
\bibinfo{author}{\bibfnamefont{B.~N.~J.} \bibnamefont{Persson}},
  \emph{\bibinfo{title}{Sliding Friction: Physical Principles and
  Applications}} (\bibinfo{publisher}{Springer-Verlag Berlin},
  \bibinfo{year}{2000}).

\bibitem[{\citenamefont{van Wijk et~al.}(2016)\citenamefont{van Wijk, de~Wijn,
  and Fasolino}}]{merel}
\bibinfo{author}{\bibfnamefont{M.~M.} \bibnamefont{van Wijk}},
  \bibinfo{author}{\bibfnamefont{A.~S.} \bibnamefont{de~Wijn}},
  \bibnamefont{and} \bibinfo{author}{\bibfnamefont{A.}~\bibnamefont{Fasolino}},
  \bibinfo{journal}{J. Phys.: Cond. Matter} \textbf{\bibinfo{volume}{28}},
  \bibinfo{pages}{134007} (\bibinfo{year}{2016}),
  \urlprefix\url{http://stacks.iop.org/0953-8984/28/i=13/a=134007}.

\bibitem[{\citenamefont{Kaya et~al.}(2013)\citenamefont{Kaya, Schlesinger,
  Yamamoto, Newberg, Bluhm, Ogasawara, Kendelewicz, Brown~Jr., Pettersson, and
  Nilsson}}]{Kaya-BaF2-2013}
\bibinfo{author}{\bibfnamefont{S.}~\bibnamefont{Kaya}},
  \bibinfo{author}{\bibfnamefont{D.}~\bibnamefont{Schlesinger}},
  \bibinfo{author}{\bibfnamefont{S.}~\bibnamefont{Yamamoto}},
  \bibinfo{author}{\bibfnamefont{J.~T.} \bibnamefont{Newberg}},
  \bibinfo{author}{\bibfnamefont{H.}~\bibnamefont{Bluhm}},
  \bibinfo{author}{\bibfnamefont{H.}~\bibnamefont{Ogasawara}},
  \bibinfo{author}{\bibfnamefont{T.}~\bibnamefont{Kendelewicz}},
  \bibinfo{author}{\bibfnamefont{G.~E.} \bibnamefont{Brown~Jr.}},
  \bibinfo{author}{\bibfnamefont{L.~G.~M.} \bibnamefont{Pettersson}},
  \bibnamefont{and} \bibinfo{author}{\bibfnamefont{A.}~\bibnamefont{Nilsson}},
  \bibinfo{journal}{Sci. Rep.} \textbf{\bibinfo{volume}{3}},
  \bibinfo{pages}{1074} (\bibinfo{year}{2013}).

\bibitem[{\citenamefont{Soper and Ricci}(2000)}]{Soper-HDL-LDL-2000}
\bibinfo{author}{\bibfnamefont{A.~K.} \bibnamefont{Soper}} \bibnamefont{and}
  \bibinfo{author}{\bibfnamefont{M.~A.} \bibnamefont{Ricci}},
  \bibinfo{journal}{Phys. Rev. Lett.} \textbf{\bibinfo{volume}{84}},
  \bibinfo{pages}{2881} (\bibinfo{year}{2000}).

\bibitem[{\citenamefont{Shiratani and Sasai}(1996)}]{Shiratani1996}
\bibinfo{author}{\bibfnamefont{E.}~\bibnamefont{Shiratani}} \bibnamefont{and}
  \bibinfo{author}{\bibfnamefont{M.}~\bibnamefont{Sasai}}, \bibinfo{journal}{J.
  Chem. Phys.} \textbf{\bibinfo{volume}{104}}, \bibinfo{pages}{7671}
  (\bibinfo{year}{1996}).

\bibitem[{\citenamefont{Shiratani and Sasai}(1998)}]{Shiratani1998}
\bibinfo{author}{\bibfnamefont{E.}~\bibnamefont{Shiratani}} \bibnamefont{and}
  \bibinfo{author}{\bibfnamefont{M.}~\bibnamefont{Sasai}}, \bibinfo{journal}{J.
  Chem. Phys.} \textbf{\bibinfo{volume}{108}}, \bibinfo{pages}{3264}
  (\bibinfo{year}{1998}).

\bibitem[{\citenamefont{Lechner and Dellago}(2008)}]{dellagobondorderparams}
\bibinfo{author}{\bibfnamefont{W.}~\bibnamefont{Lechner}} \bibnamefont{and}
  \bibinfo{author}{\bibfnamefont{C.}~\bibnamefont{Dellago}},
  \bibinfo{journal}{J. Chem. Phys.} \textbf{\bibinfo{volume}{129}},
  \bibinfo{eid}{114707} (\bibinfo{year}{2008}),
  \urlprefix\url{http://scitation.aip.org/content/aip/journal/jcp/129/11/10.1063/1.2977970}.

\bibitem[{\citenamefont{Steinhardt et~al.}(1983)\citenamefont{Steinhardt,
  Nelson, and Ronchetti}}]{steinhardtbondorderparams}
\bibinfo{author}{\bibfnamefont{P.~J.} \bibnamefont{Steinhardt}},
  \bibinfo{author}{\bibfnamefont{D.~R.} \bibnamefont{Nelson}},
  \bibnamefont{and}
  \bibinfo{author}{\bibfnamefont{M.}~\bibnamefont{Ronchetti}},
  \bibinfo{journal}{Phys. Rev. B} \textbf{\bibinfo{volume}{28}},
  \bibinfo{pages}{784} (\bibinfo{year}{1983}),
  \urlprefix\url{http://link.aps.org/doi/10.1103/PhysRevB.28.784}.

\bibitem[{\citenamefont{Kamb and Davis}(1964)}]{iceVII}
\bibinfo{author}{\bibfnamefont{B.}~\bibnamefont{Kamb}} \bibnamefont{and}
  \bibinfo{author}{\bibfnamefont{B.~L.} \bibnamefont{Davis}},
  \bibinfo{journal}{Proc. Nat. Acad. Sci. USA} \textbf{\bibinfo{volume}{52}},
  \bibinfo{pages}{1433} (\bibinfo{year}{1964}).

\bibitem[{\citenamefont{Peyrard and Aubry}(1983)}]{vanishingstaticfriction}
\bibinfo{author}{\bibfnamefont{M.}~\bibnamefont{Peyrard}} \bibnamefont{and}
  \bibinfo{author}{\bibfnamefont{S.}~\bibnamefont{Aubry}},
  \bibinfo{journal}{J.~Phys.~C} \textbf{\bibinfo{volume}{16}},
  \bibinfo{pages}{1593} (\bibinfo{year}{1983}).

\bibitem[{\citenamefont{M\"user et~al.}(2001)\citenamefont{M\"user, Wenning,
  and Robbins}}]{mueserprl}
\bibinfo{author}{\bibfnamefont{M.~H.} \bibnamefont{M\"user}},
  \bibinfo{author}{\bibfnamefont{L.}~\bibnamefont{Wenning}}, \bibnamefont{and}
  \bibinfo{author}{\bibfnamefont{M.~C.} \bibnamefont{Robbins}},
  \bibinfo{journal}{Phys. Rev. Lett.} \textbf{\bibinfo{volume}{86}},
  \bibinfo{pages}{1295} (\bibinfo{year}{2001}).

\bibitem[{\citenamefont{de~Wijn}(2012)}]{astridgoldgraphite}
\bibinfo{author}{\bibfnamefont{A.~S.} \bibnamefont{de~Wijn}},
  \bibinfo{journal}{Phys. Rev. B} \textbf{\bibinfo{volume}{86}},
  \bibinfo{pages}{085429} (\bibinfo{year}{2012}).

\bibitem[{\citenamefont{Dietzel et~al.}(2013)\citenamefont{Dietzel, Feldmann,
  Schwarz, Fuchs, and Schirmeisen}}]{dietzelscaling}
\bibinfo{author}{\bibfnamefont{D.}~\bibnamefont{Dietzel}},
  \bibinfo{author}{\bibfnamefont{M.}~\bibnamefont{Feldmann}},
  \bibinfo{author}{\bibfnamefont{U.~D.} \bibnamefont{Schwarz}},
  \bibinfo{author}{\bibfnamefont{H.}~\bibnamefont{Fuchs}}, \bibnamefont{and}
  \bibinfo{author}{\bibfnamefont{A.}~\bibnamefont{Schirmeisen}},
  \bibinfo{journal}{Phys. Rev. Lett.} \textbf{\bibinfo{volume}{111}},
  \bibinfo{pages}{235502} (\bibinfo{year}{2013}).

\bibitem[{\citenamefont{Klotz et~al.}(1999)\citenamefont{Klotz, Besson, Hamel,
  Nelmes, Loveday, and Marshall}}]{metastableiceVII}
\bibinfo{author}{\bibfnamefont{S.}~\bibnamefont{Klotz}},
  \bibinfo{author}{\bibfnamefont{J.~M.} \bibnamefont{Besson}},
  \bibinfo{author}{\bibfnamefont{G.}~\bibnamefont{Hamel}},
  \bibinfo{author}{\bibfnamefont{R.~J.} \bibnamefont{Nelmes}},
  \bibinfo{author}{\bibfnamefont{J.~S.} \bibnamefont{Loveday}},
  \bibnamefont{and} \bibinfo{author}{\bibfnamefont{W.~G.}
  \bibnamefont{Marshall}}, \bibinfo{journal}{Nature}
  \textbf{\bibinfo{volume}{398}}, \bibinfo{pages}{681} (\bibinfo{year}{1999}).

\end{thebibliography}
\end{document}